\documentclass[%
 reprint,
 jmp,
superscriptaddress,
 amsmath,amssymb,
 aps,
]{revtex4-2}

\usepackage{graphicx}
\usepackage{dcolumn}
\usepackage{bm}
\usepackage{svg}

\usepackage{xcolor}
\begin{document}

\preprint{APS/123-QED}

\title{Uncorrelated photon pair generation from an integrated silicon nitride resonator measured by time resolved coincidence detection}
\author{Massimo Borghi}
\email{corresponding author: massimo.borghi@unipv.it} 
\affiliation{Dipartimento di Fisica, Università di Pavia, Via Bassi 6, 27100 Pavia, Italy.}
\author{Paula L. Pagano}
\affiliation{
Centro de Investigaciones Ópticas, CONICET-CICBA-UNLP, Camino Centenario y 505 s/n , 1897 Gonnet, Argentina.}
\affiliation{Dipartimento di Ingegneria Industriale e dell'Informazione, Università di Pavia, Via Ferrata 5, 27100 Pavia, Italy.}


\author{Marco Liscidini}
\affiliation{Dipartimento di Fisica, Università di Pavia, Via Bassi 6, 27100 Pavia, Italy.}

\author{Daniele Bajoni}
\affiliation{Dipartimento di Ingegneria Industriale e dell'Informazione, Università di Pavia, Via Ferrata 5, 27100 Pavia, Italy.}

\author{Matteo Galli}
\affiliation{Dipartimento di Fisica, Università di Pavia, Via Bassi 6, 27100 Pavia, Italy.}

\date{\today}

\begin{abstract}
We measure the joint temporal intensity of signal and idler photon pairs generated by spontaneous four wave mixing in a silicon nitride microresonator by time-resolved coincidence detection. This technique can be applied to any high-Q optical cavity whose photon lifetime exceeds the duration of the pump pulse. We tailor the temporal correlation of photon pairs by using a resonant interferometric coupler, a device that allows us to independently tune the quality factors of the pump and signal and idler resonances.
Temporal post-selection is used to accurately measure the temporal emission of the device, demonstrating a purity of $98.67(1)\%$.
\end{abstract}

\maketitle


\section{\label{sec:intro} Introduction}
Photon pairs generated by spontaneous four wave mixing (SFWM) or parametric down conversion (PDC) play a key role in quantum information processing. Time-energy entangled photons find use in unconditionally secure communication \cite{joshi2020trusted}, in sensing beyond the standard quantum limit \cite{zhang2021distributed} and in quantum-enhanced illumination \cite{gregory2020imaging}. Single photons can be heralded and used for quantum computation \cite{alexander2024manufacturable} and simulation \cite{paesani2019generation}. 
    In all these applications, energy correlations of the generated photons must be controlled and characterized. These are usually described in terms of the joint spectral(temporal) amplitude (resp. JSA and JTA), whose modulus square represents the probability density of detecting the two photons in frequency(time) \cite{quesada2022beyond}. \\
Several techniques have been developed to measure the JTA and JSA \cite{gianani2020measuring}. When the spectral emission is broadband 
frequency-resolved coincidence detection is more practical than time-resolved detection due to the availability of narrow-band spectral filters with GHz resolution \cite{faruque2023quantum}. Stimulated emission tomography 
is an alternative solution that drastically reduces the acquisition time by increasing the strength of the generated signal by orders of magnitude  \cite{liscidini2013stimulated,grassani2016energy}. 
In optical cavities with sub-GHz linewidths \cite{huang2023integrated}, or for SFWM photons generated in atomic clouds where the emission bandwidth is well below the MHz regime \cite{
park2017measuring}, narrow-band filtering is very challenging and more complex strategies are required \cite{jeong2020joint}. For photons with coherence times exceeding hundreds of ps, time-resolved coincidence measurements performed with commercial single photon detectors provide sufficient resolution to reveal fine-grain temporal correlations to quantify the amount of time-energy entanglement \cite{yang2018tomography}. While this strategy is of widespread use for the characterization of photon pairs of ultra-narrow bandwidth generated by atomic vapors, its implementation in high-Q microcavities is far less common. Partially, this is because the heralded single-photon purity, can also be addressed from the measurement of the unheralded second-order correlation ($g_2$) of the signal or the idler arm \cite{christ2011probing}. However, the heralded single-photon purity can be extracted from the JSA(JTA) with much higher accuracy, because the coincidence detection is less prone to noise from uncorrelated photons in either the signal or the idler arms \cite{eckstein2011realistic}. The JSA(JTA) is more informative since it allows us to infer the distribution of the Schmidt modes, which is relevant for the characterization of photonic qubits encoded in pulsed-temporal modes \cite{ansari2018tailoring}. \\
In this work we implement time-resolved coincidence measurements to retrieve the joint temporal intensity (JTI) of photon pairs generated by SFWM in an integrated silicon nitride resonator.  \color{black} These devices are used in integrated quantum optical processors for the generation of bright squeezed light, and take advantage of the unique combination of low propagation losses, high nonlinearity, lack of two-photon absorption and mature fabrication process offered by the silicon nitride platform \cite{alexander2024manufacturable,arrazola2021quantum}. 
We use a resonant interferometric coupler structure, introduced in \cite{pagano2024selective}, to selectively change the Q-factor of 
the pump resonance and tailor the time correlation of the photon pairs \cite{vernon2017truly}. \color{black} From the JTI we extracted a maximum purity of $0.9867(1)$, lying well above the upper bound of $0.93$ for a resonator with equal quality factors \cite{vernon2017truly}. Finally, we compare the value of the purity extracted from the temporally post-selected JTI to that from the unheralded $g_2$.  \color{black}The goal of this work is to present a technique which allows to measure the purity of photon pairs with high accuracy, which is robust against background noise and that can be applied to spectrally narrow-band sources.\color{black}
\section{Device and experimental setup}
\begin{figure}[h!]
    \centering
    \includegraphics[width = 0.48\textwidth]{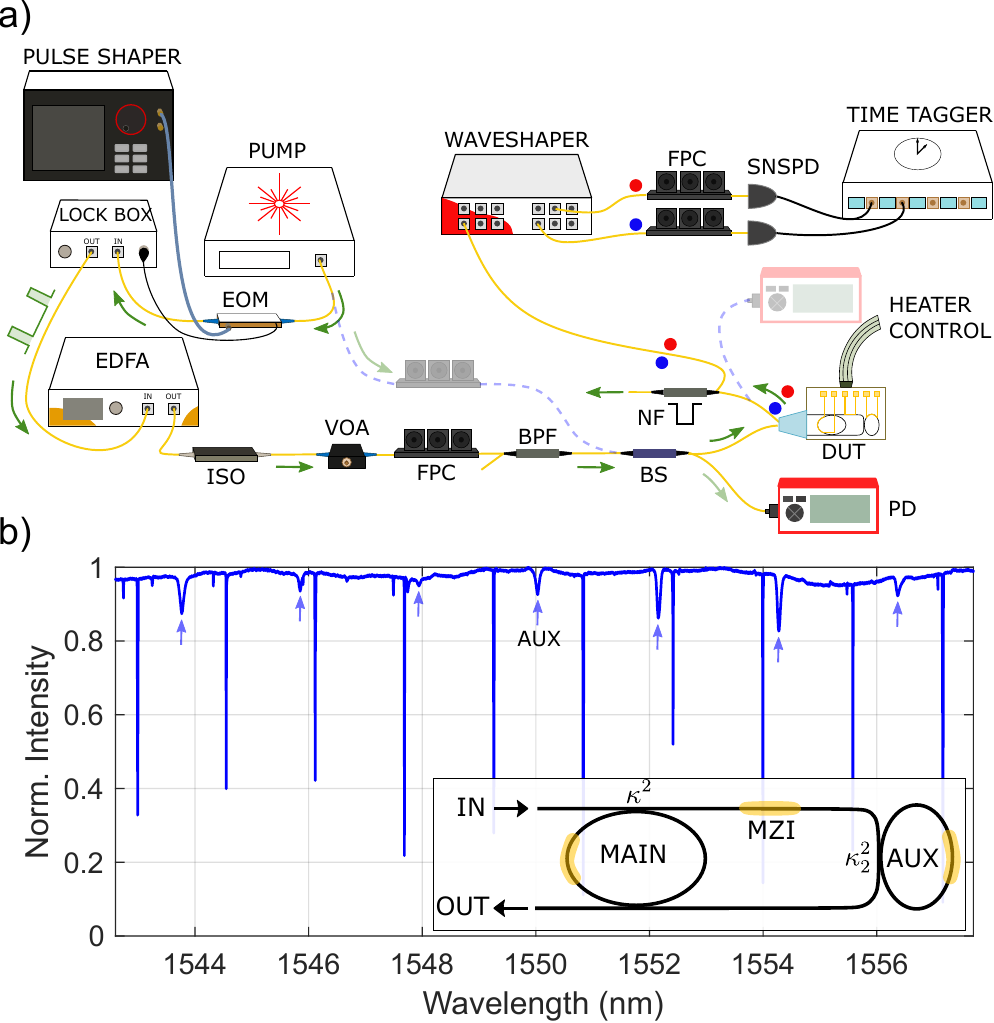}
    \caption{(a) Sketch of the experimental setup.  \color{black} The dashed blue lines are the optical connections used to probe the spectral response of the device. \color{black} EOM: electro-optic amplitude modulator, EDFA: erbium doped fiber amplifier, ISO: isolator, VOA: variable optical attenuator, FPC: fiber polarization controller, BS: fiber beam splitter, PD: photodiode, DUT: device under test, NF: notch filter, SNSPD: superconducting nanowire single photon detector. (b) Spectral response of the device. Blue arrows indicate the resonances of the Aux. resonator. The inset shows a sketch of the device. 
    Thermal phase shifters are shown in yellow.}
    \label{Fig_1}
\end{figure}

Our device is fabricated on a silicon nitride photonic chip. A sketch is shown in the inset of Fig.\ref{Fig_1}(b). It consists of a main resonator of perimeter $L_m$ that is coupled twice to a U-shaped bus waveguide of length $L_{\textup{MZI}}=2L_m$, forming an asymmetric Mach-Zehnder interferometer (MZI) \cite{vernon2017truly}.  \color{black} The geometrical parameters and the linear characterization of the device can be found in Appendix A and in \cite{pagano2024selective}. \color{black} By changing the phase $\theta_{\textup{MZI}}$ inside the MZI through a thermo-optic phase shifter, the effective coupling between the main resonator and the waveguide can be continuously tuned from zero to $4\kappa^2(1-\kappa^2)$, in which $\kappa^2$ is the coupling coefficient of the two point couplers of the main resonator. 
Our choice of $L_{\textup{MZI}}$ ensures that each resonance experiences the same value of the effective coupling coefficient \cite{pagano2024selective}. An auxiliary (Aux.) resonator of perimeter $L_a = (3/4)L_m$ is overcoupled ($\kappa_2 = 0.21$) to the external arm of the MZI. The Aux. resonator is used to impart a resonant additional phase $\theta_{\textup{aux}}$ inside the MZI. This phase is negligible at the wavelengths lying far from any resonance, but abruptly changes from $\theta_{\textup{aux}} = 0$ to $\theta_{\textup{aux}} = 2\pi$ at  resonance \cite{bogaerts2012silicon}. 
Thus the resonant interferometric coupler allows us to selectively change the coupling rate (and thus the quality factor) of a single resonance of the main resonator with minimum perturbation of the neighboring ones. \\
We use a simplified version  \color{black}(optical connections indicated with dashed blue lines) \color{black} of the experimental setup shown in Fig.\ref{Fig_1}(a)  to probe the spectral response of the device. Light from a tunable continous wave laser is passed through a fiber polarization controller and coupled to the input waveguide of the photonic chip by means of an array of UHNA4 fibers. The output is collected from the same array and monitored by a photodiode while the input wavelength is continously scanned. An example of a recorded spectra is shown in Fig.\ref{Fig_1}(b). We identify the comb of resonances of the main resonator as narrow dips with a spacing of $\sim 200 \,\textrm{GHz}$ ($1.56$ nm). In this case, the phase $\theta_{\textup{MZI}}$ is configured to set the Q-factor to $\sim 8\times10^5$. 
The comb of resonances of the Aux. resonator appear as a family of dips with low extinction ($\sim-0.3$ dB) and a spacing of $\sim 266\,\textrm{GHz}$ ($2.08$ nm). The Q-factor of the Aux. resonances is $\sim 2.6\times 10^4$.\\
One of the resonances of the main resonator is pumped by the input laser, triggering the generation of signal-idler photon pairs by SFWM. 
We carve rectangular pump pulses of duration of $T=300$ ps and a repetition rate of $10$ MHz by passing the continous wave laser into an electro-optic amplitude modulator (Fig.\ref{Fig_1}(a)). 
The generated signal-idler pairs are filtered from the pump by using an off-chip notch filter, and demultiplexed to two separate superconducting nanowire single photon detectors (SNSPD). The outputs of the SNSPDs are connected to a time-tagging unit with a timing-jitter of $35$ ps. The unit is synchronized with the electrical trigger of the pump pulse to retrieve the arrival times of the two photons relative to the trigger. 

\section{Measurement of the joint temporal intensity and temporal post-selection}
\begin{figure}[t!]
    \centering
    \includegraphics[width = 0.47\textwidth]{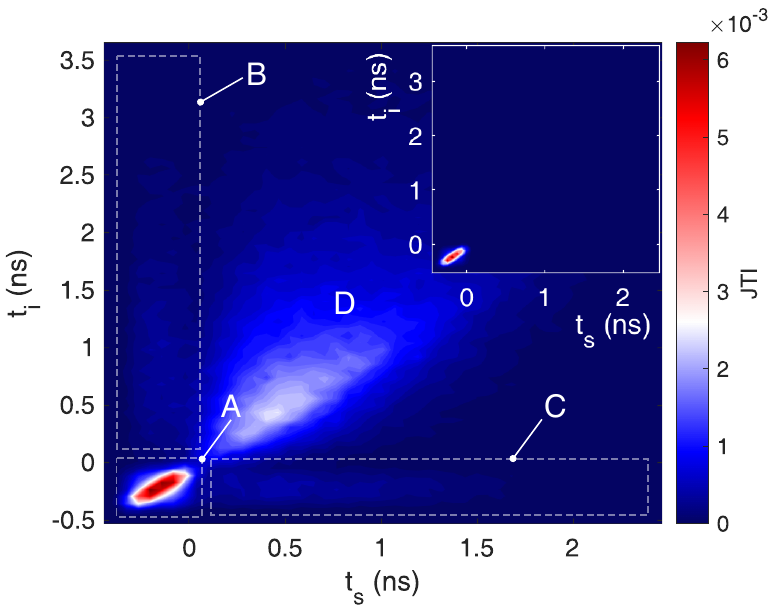}
    \caption{Example of a measured JTI. The four regions enclosed by the dashed boxes, labeled from A to D, collect coincidences arising from distinct SFWM processes, described in the main text. The top right inset shows the JTI when the pump is not resonant with the main resonator.} 
    \label{Fig_2}
\end{figure}

In a preliminary measurement, $\theta_{\textup{MZI}}$ is configured to set the Q-factor of the pump at $\lambda_p=1543$ nm to $Q_p\sim10^6$, and the Aux. resonator is used to set the Q-factor of the signal resonance at $\lambda_s = 1541.4$ nm to $Q_s\sim2.7\times 10^5$. The idler resonance  \color{black} at $1544.6$ nm \color{black} has a Q-factor of $Q_i\sim 8\times10^5$. The asymmetry between $Q_p$ and $Q_i$ arises from the presence of an Aux. resonance in neighborhood of a pump resonance. The arrival times $t_s$ and $t_i$ of the signal and the idler photon are recorded for $10$ minutes and subsequently binned on a two dimensional grid with a resolution of $\sqrt{3}\times35\sim60$ ps, which is determined by convolving the timing jitter of the channels involved in the measurement record. From the two dimensional histogram we retrieve the joint temporal intensity JTI($t_s,t_i$), describing the joint probability of a signal photon arriving at time $t_s$ and an idler photon at time $t_i$. This is shown in Fig.\ref{Fig_2}. A rich structure is revealed, corresponding to different physical processes occurring in the device, and that we group into four distinct regions labeled from A to D. The majority of signal-idler photons arrive to the detectors in a time window of approximately $300$ ps (region A), and their arrival times are positively correlated. The time window is comparable to the pump pulse duration $T$, which suggests that they may be generated by SFWM in the bus waveguide. To check this hypothesis, we repeated the measurement with the pump laser out of resonance, and we obtained the JTI shown in the inset of Fig.\ref{Fig_2}. 
The coincidences are now all clustered in region A, while the events in regions B, C and D disappeared. This confirms that photon pairs arriving in region A are indeed generated in the bus waveguide. 
We can now identify the different mechanisms leading to coincidence detections in regions B, C and D. In D, both photons are generated within the main resonator. 
Once the pump pulse has been loaded into the cavity, the pump photons circulate in the resonator for about  $\tau_p=\frac{Q_p}{\omega_p}\sim820$ ps, during which they trigger the generation of pairs by SFWM, which is shown in Fig.\ref{Fig_2}. Interestingly, region A and D do not overlap, i.e., photons generated from the waveguide can be temporally distinguished from those generated by the resonator. This happens because $\tau_p>T$, implying that the time required for the pump energy to build up inside the resonator and trigger SFWM exceeds the temporal pulse width during which pairs are generated in the bus waveguide. Within this framework, coincidences collected in region B correspond to emission of double pairs, in which a signal from the waveguide is followed by a detection of an idler from the resonator. Finally, events in region C corresponds to the generation of double pairs where an idler from the waveguide is followed by a detection of a signal from the resonator. It is worth to stress that in the frequency domain, the JSI associated with photon pairs generated in the waveguide and in the resonator overlap. If these contributions do not coherently interfere because they can be distinguished in time, as in our case, the two JSIs add up incoherently, and one has to perform two measurement records (on and off resonance) to retrieve the JSI of the resonator. 
In our system we can tailor the quality factor of the resonances involved in SFWM, thus changing the energy correlation of the generated pairs.
In what follows, all the reported JTIs are reconstructed from the coincidences collected in region D of Fig.\ref{Fig_2} for different configurations. In the first one, the pump resonance lies in between two consecutive resonances of the Aux. resonator to achieve minimal perturbation. 
\begin{figure*}[t!]
    \centering
    \includegraphics[width = 0.95\textwidth]{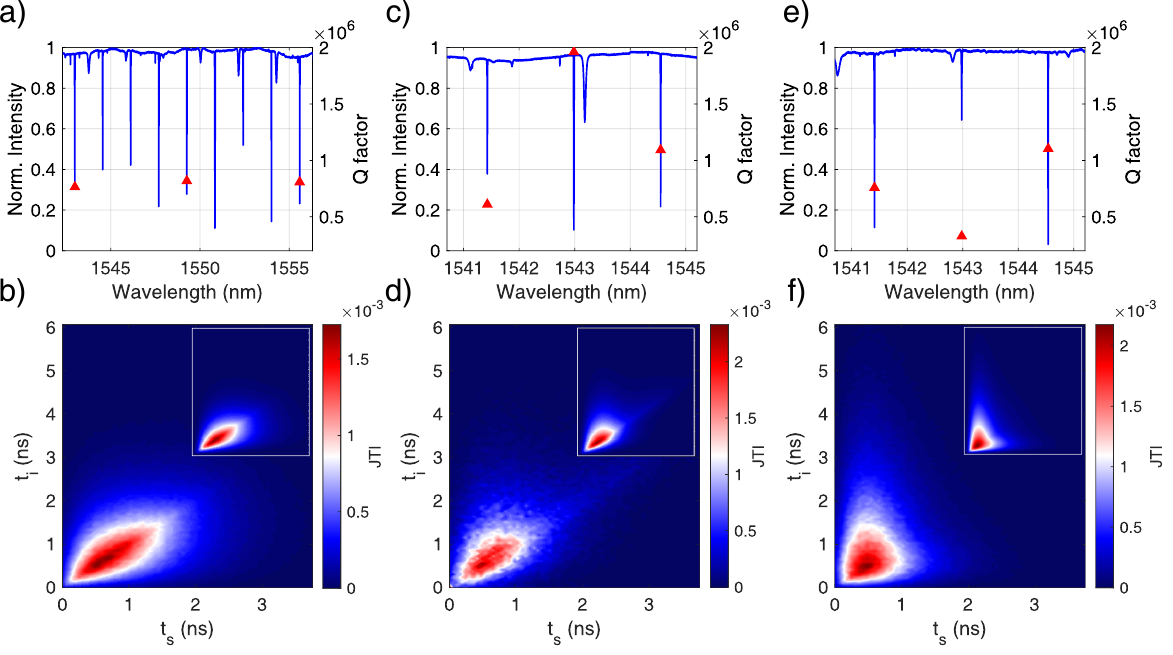}
    \caption{Transmission spectra of the device for three different settings of $\theta_{\textup{MZI}}$ and $\theta_{\textup{aux}}$ and measured JTI. In the reference configuration (a,b) $Q_p\sim Q_s\sim Q_i$. In (c,d) $Q_p\gg (Q_s,Q_i)$, while in (e,f) $Q_p<(Q_s,Q_i)$. In panels (a,c,d) the Q-factor of the signal, pump and idler resonance is indicated with a red triangle. The inset in the top right corner of panels (b,d,f) shows the simulated JTI.}
    \label{Fig_3}
\end{figure*}
As shown in Fig.\ref{Fig_3}(a), we choose the signal and the idler resonances lying at four FSRs from the pump wavelength. The periodicity imposed by the choice of $L_\textup{aux}$ ensures that this set of resonances have the same $Q=8\times10^5$. This configuration mimics what happens in a conventional resonator that is point coupled to a bus waveguide \cite{bogaerts2012silicon}. When the Q-factors of all the three resonances involved in the SFWM process are equal and the pump shape is gaussian, the spectral purity of the heralded photons is limited to $0.93$ \cite{vernon2017truly}. The reconstructed JTI is shown in Fig.\ref{Fig_3}(b). 
Following \cite{vernon2017truly}, from the measured JTI we estimate an upper bound to the heralded photon purity of $0.9289(2)$, corresponding to a Schmidt number of $K=1.0765(4)$. This value is very close to theoretical upper bound. 
In the second configuration, shown in Fig.\ref{Fig_3}(c), the Aux. resonator is used to selectively tune the Q-factor of the pump resonance to \mbox{$Q_p = 1.96\times10^6$}, while the signal and the idler resonances have a quality factor of respectively $Q_s =6.1\times10^5$ and $Q_i = 1\times10^6$.  The asymmetry between $Q_s$ and $Q_i$ follows from the proximity of the signal resonance to an Aux. resonance. The measured JTI is shown in Fig.\ref{Fig_3}(d). The choice of a higher Q-factor for the pump compared to that of the signal and the idler results in a JTI that is  \color{black} slightly more elongated \color{black} along the main anti-diagonal, enhancing temporal correlations. As expected, the extracted purity is $0.799(2)$ ($K=1.251(3)$), significantly lower than the value of the reference configuration. To check the reliability of the reconstruction, we simulated the JTI using the method of asymptotic fields described in \cite{liscidini2012asymptotic}, where the field distribution in the device is obtained by the methods described in  \cite{pagano2024selective}. The result of the simulation, shown in the inset of Fig.\ref{Fig_3}(d), well agrees with the measurement. The purity calculated from the simulated JTI is $0.829$ ($K=1.20$). The purity calculated from the complex JTA is $0.548$, indicating the presence of strong phase correlations that are hidden in the JTI.\\
In the third configuration, shown in Fig.\ref{Fig_3}(e), the Aux. resonator is used to set the Q-factor of the pump to $Q_p=3.3\times10^5$, while $Q_s=7.6\times10^5$ and $Q_i=1.1\times10^6$. As demonstrated in \cite{vernon2017truly}, when the Q-factor of the pump is lower than that of the signal and the idler, photon pairs can be generated with a purity beyond $0.93$. In our case, $\frac{\textrm{min}(Q_s,Q_i)}{Q_p}\sim 2.3>1$. We report the reconstructed JTI in Fig.\ref{Fig_3}(f). Correlations between $t_s$ and $t_i$ are now greatly suppressed, and the purity is $0.9867(1)$ ($K=1.0134(2)$), which indicates that the emission is almost single mode. The simulated JTI, shown in the inset of Fig.\ref{Fig_3}(f), has a purity of $0.985$, while the value calculated from the complex JTA is $0.9847$, indicating negligible phase correlations. This value is among the highest ever reported for an integrated resonator  \color{black}(see Appendix B for a comparison with similar works).\color{black} \\ 
The JTI allows us to estimate the Schmidt number in high Q resonators with better accuracy compared to the more conventional measurement of the time-integrated unheralded second-order correlation ($g_2$) on the signal or the idler beam \cite{christ2011probing}. This is because noise photons 
alter the genuine photon statistics of the unheralded arm. 
To further highlight this point, we measured the integrated $g_2$ in the configuration shown in Fig.\ref{Fig_3}(e), obtaining $g_2=1.70(2)$ and a corresponding purity of $g_2-1=0.70(2)$, a value that is significantly lower than that extracted from the JTI. 
To validate the impact of noise photons on the value of the $g_2$, we measured how the signal(idler) click probability $p_{s(i)}$ scale as a function of the pump power $P$, and fit the data with the model $N_{s(i)}  = \sum_{n=1}^{\infty}\eta_{s(i)}^n\rho_n^{(s,i)}$,
where $\eta_{s(i)}$ is the signal(idler) loss from the chip to the detector 
and $\rho_{n}^{(s,i)}$ is the photon number distribution of the signal(idler) mode. The latter is modeled as the convolution of two uncorrelated thermal states $\hat{\rho}_{\textup{th,1(2)}}$ with mean photon number $\langle n_1\rangle=\sinh^2{(aP)}$ and $\langle n_{2,s(i)}\rangle = b_{s(i)}P$, where $a$ and $b_{s(i)}$ are constants that are determined from the fit. The term $\hat{\rho}_{\textup{th,1}}$ describes the state of the signal(idler) after that the idler(signal) has been traced out from the two-mode squeezed state, while $\hat{\rho}_{\textup{th,2}}$ accounts for any thermal noise contribution. At the pump power used for of the $g_2$ measurement, we extracted $\langle n_1\rangle = 0.21(3)$, $\langle n_{2,s} \rangle = 0.042(2)$ and \mbox{$\langle n_{2,i} \rangle = 0.048(3)$}.  \color{black} The brightness of our source is $2.5\times10^{-6}\,\frac{\langle n \rangle}{\textrm{pJ}^2}$ and the intrinsic (preparation) heralding efficiency is $\eta_{\textup{HE}}=0.23(2)$. \color{black} The calculated $g_2$ from these parameters is $g_2 = 1.73(2)$ and is consistent with the experiment.  

           

\section{Conclusions}
We implemented time resolved coincidence detection to measure the joint temporal intensity of photon pairs generated by a silicon nitride microresonator. We use a resonant interferometric coupler to tailor the spectral correlations between the signal and the idler photons by selectively changing the Q-factor of the pump resonance. We measured a lower bound of the Schmidt number of $K=1.0134(2)$. The use of time resolved detection allows us to discriminate the photon pairs generated from the resonator from spurious pairs generated within the photonic circuit, thus improving the estimation of the correlations. 
This can be applied to all situations where the pump pulse duration is smaller than the dwelling time of photons in the cavity, which is a stringent requirement for the generation of unentangled pairs. 
The method is relevant for high Q microcavities of ultra-narrow linewidth, where frequency resolved detection is challenging.
\subsection*{Funding and acknowledgements}
P.L. Pagano acknowledges funding from Hyper–Space project (Project ID No. 101070168), co–funded by the EU and the Natural Sciences and Engineering Research Council of Canada (NSERC).
D.B acknowledges the support of Italian MUR and the European Union - Next Generation EU through the PRIN project number F53D23000550006 - SIGNED. M.B., M.G and M.L. acknowledge the PNRR MUR project PE0000023-NQSTI. All the authors acknowledge the support of Xanadu Quantum Technology for providing the samples and the useful discussions.

\section*{Appendix A: Device geometry}
\label{Sec:1}
\begin{figure}[h!]
    \centering
    \includegraphics[width = 0.47\textwidth]{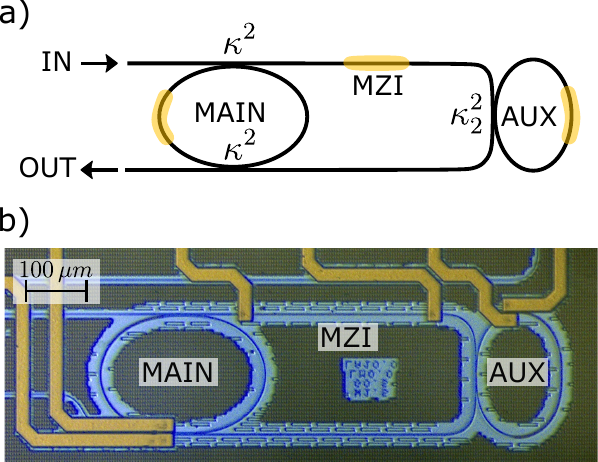}
    \caption{(a) Sketch of the device. Heaters are shown in yellow. The coupling coefficient between the main ring and the bus waveguide is $\kappa^2$, while that between the auxiliary ring and the bus waveguide is $\kappa_2^2$.}
    \label{Fig_1S}
\end{figure}
The device is fabricated on a silicon nitride photonic chip in a standard multi-project wafer run by Ligentech. A sketch is shown in Fig.\ref{Fig_1S}(a), while an optical microscope image is shown in Fig.\ref{Fig_1S}(b). All the waveguides have width of \mbox{1.75 $\mu$m} and a thickness of \mbox{800 nm}, and are covered by a silicon dioxide cladding of thickness $4.1\,\mu$m. The gap between the main resonator and the bus-waveguide is \mbox{0.4 $\mu$m}, and the corresponding coupling coefficient is $\kappa^2=0.03$. The coupling between the bus-waveguide and the auxiliary ring is achieved through a directional coupler with a length of $10\,\mu$m and a gap of \mbox{0.4 $\mu$m}. The corresponding coupling coefficient is $\kappa_2^2=0.2$. Both the main and the aux. resonators are made by two U-shaped Euler bends, which are designed to minimize the bending loss, thus maximizing the intrinsic quality factor of the devices. The linear propagation losses of the Euler have been found comparable to those of the straigth waveguide sections, which are $\sim0.11$ dB/cm. Three metallic heaters of Titanium Nitride of $0.4\,\mu$m thickness and $2\,\mu$m width are placed $1.7\,\mu$m above the waveguides. They allow to respectively change the resonance wavelength of the main ring, of the auxiliary resonator, and the phase of the bus-waveguide arm (MZI in Fig.\ref{Fig_1}) through the thermo-optic effect.
\section*{Appendix B: Comparison of the performance metrics with other microresonator-based sources}
\label{Sec:2}
In Table \ref{tab:1} we compare some of the most relevant metrics of our device with similar works using microresonators as souces of heralded single-photons. \\
We choose to focus on the brightness, the intrinsic (preparation) heralding efficiency and the purity. The brightness is expressed as the average number of pairs generated per pulse $\Bar{n}$ per square of the pulse energy (expressed in $\textrm{pJ}^2$). This choice allows us to remove the dependence of the brightness on both the pulse duration and the laser repetition rate, and allows us to fairly compare our results to those reported in other similar works. \\
The intrinsic heralding efficiency $\eta_{\textup{HE}}$ is defined as the probability that, given the detection of an herald photon, the heralded arm contains a photon in the limit of zero loss. We can calculate $\eta_{\textup{HE}}$ from the following expression 
\begin{equation}
    \eta_{\textup{HE}} = \frac{R_{si}}{R_s\eta_i}, \label{eq:1}
\end{equation}
where $R_{si}$ is the coincidence rate at the detectors, $R_s$ is the rate of herald photons, $\eta_i=\eta_d T_i$, $T_i$ is the transmittivity of the heralded arm and $\eta_d$ is the detection efficiency. In our experiment, we measured $\eta_d=0.85$ and $T_i=0.073$. 
\begin{table}[b!]
\centering
\caption{\bf Comparison of the main performance metrics between our work and other microresonator-based heralded single-photon sources}
\begin{tabular}{cccc}
\hline
Reference & Brightness $\left (\frac{\bar{n}}{\textrm{pJ}^2}\right )$ & Purity & Heralding efficiency $\eta_{\textup{HE}}$ \\
\hline
\cite{liu2020high} & $6.9\times10^{-5}$ & $0.95$ & $0.524$ \\
\cite{burridge2023integrate} & $2.24\times10^{-4}$ & $0.991$ & $0.94$ \\
\cite{silverstone2015qubit} & $10^{-2}$ & $0.855$ & $0.5$ \\
\cite{savanier2016photon} & $3.6\times10^{-2}$ & $0.917$ & $0.5$ \\
\cite{burridge2020high} & $1.7\times10^{-7}$ & $0.98$ & $0.5$ \\
This work & $2.5\times10^{-6}$ & $0.9867$ & $0.23$ \\
\hline
\end{tabular}
  \label{tab:1}
\end{table}

\providecommand{\noopsort}[1]{}\providecommand{\singleletter}[1]{#1}%

\end{document}